\documentclass[preprint,showpacs,preprintnumbers,amsmath,amssymb]{revtex4}

\usepackage{graphicx,subfigure}
\usepackage{dcolumn}
\usepackage{bm}
\usepackage{amsmath}

\newcommand{\be}{\begin{equation}}
\newcommand{\ee}{\end{equation}}
\newcommand{\bea}{\begin{eqnarray}}
\newcommand{\eea}{\end{eqnarray}}

\begin{document}

\bibliographystyle{apsrev}

\preprint{UAB-FT-577}

\title{Scalar Field Oscillations Contributing \\
to Dark Energy
} 

\author{Eduard Mass{\'o}}

\author{Francesc Rota}

\author{Gabriel Zsembinszki}

\affiliation{Grup de F{\'\i}sica Te{\`o}rica and Institut
de F{\'\i}sica d'Altes
Energies\\Universitat Aut{\`o}noma de Barcelona\\
08193 Bellaterra, Barcelona, Spain}


\date{\today}

\begin{abstract}
We use action-angle variables to describe the basic physics of
coherent scalar field oscillations in the expanding universe.
These analytical mechanics methods have some advantages, like the
identification of adiabatic invariants. As an application, we show
some instances of potentials leading to equations of state with
$p<-\rho/3$, thus contributing to the dark energy that causes the
observed acceleration of the universe.
\end{abstract}

\maketitle


\section{Introduction}\label{section1}
\label{introduction} The standard model of cosmology assumes the
Friedmann-Robertson-Walker metric
\begin{equation}
ds^2=-dt^2+R^2(t)\left[\frac{dr^2}{1-kr^2}+r^2d\Omega^2\right]
\label{FRW_metric}
\end{equation}
corresponding to an homogeneous and isotropic universe. Here $k$
is the curvature signature and $R$ is the expansion factor, whose
time change is given by the Friedmann equation
\begin{equation}
H^2\equiv\left(\frac{\dot{R}}R\right)^2=\frac{8\pi
G}3\rho_T-\frac{k}{R^2}. \label{Fried_eq}
\end{equation}
We have written the equation in such a way that the cosmological
constant is included in the total energy density $\rho_T$.

There are different contributions to $\rho_T$. Matter and
radiation are among them. They can be introduced as a fluid with
pressure proportional to the energy density, $p=w\rho$; $w=0$
corresponds to matter and $w=1/3$ to radiation. Recent results
coming from high-redshift supernovae \cite{supernova}, cosmic
background radiation \cite{cmb}, and galaxy survey \cite{galaxy}
suggest contributions with $w\approx -1$. This is the so-called
dark energy that leads to the present acceleration of the
universe. This acceleration may be produced by a cosmological
constant, since it has an equation of state with $w=-1$. However,
there might be other causes, like quintessence
\cite{quintessence}, cardassian expansion \cite{cardassian}, etc.
Of course, there might be several components building up dark
energy.

In this paper, we would like to discuss some aspects of the
contribution to the energy density of coherent scalar field
oscillations in a potential. The physics of such oscillations was
analyzed in a pioneer work by Turner \cite{Turner}. The purpose of
our paper is the following. First, in Sect.\ref{section2}, we find
the analytical form of the field oscillations in a couple of
interesting cases. In Sect.\ref{section3}, we treat the general
case and derive relevant physical results using the action and
angle variables of analytical mechanics \cite{goldstein}. We find
this language very useful, particularly when an adiabatic change
of the potential is present. Another objective of our paper is to
show that scalar field oscillations may contribute to the dark
energy of the universe. We do this in Sect.\ref{section4}, where
we present some instances of potentials leading to an accelerated
universe.

\section{Analytical solutions for scalar field oscillations}\label{section2}
A classical scalar field in a potential has the Lagrangian density
\begin{equation}
{\cal L}=-\frac12\partial_{\mu}\phi\,\partial^{\mu}\phi-V(\phi).
\end{equation}
In the metric (\ref{FRW_metric}), and considering spatially
homogeneous configurations, the equation of motion for the field
is
\begin{equation}
\ddot{\phi}+3H\dot{\phi}+V'(\phi)=0. \label{evoleq}
\end{equation}
Here dot means time-derivative and prime means $d/d\phi$.

In this section we concentrate on two particular potentials,
$V\sim\phi^2$ and $V\sim\phi^4$, where one can find analytical
solutions. While the solution for the harmonic potential is in the
literature \cite{threepapers}, the solution we find for $\phi^4$
is new, as far as we know.

For the quadratic potential
\begin{equation}
V=\frac12m^2\phi^2 \label{pot_quadrat}
\end{equation}
we have harmonic oscillations, $\phi(t)=A\sin(mt+\varphi)$, in the
case $H=0$.

When $H\neq 0$, but in the case $H\ll m$ and $\dot H/H\ll m$, we
expect a time dependent amplitude and, perhaps, a different
frequency. We make the following ansatz
\begin{equation}
\widetilde\phi(t)=A(t)\sin(\lambda mt+\varphi).\label{solphi2H1}
\end{equation}
We find $\lambda=1$ and the following equation
\begin{equation}
\dot A=-\frac32HA. \label{eqA2}
\end{equation}
The energy density $\rho$ corresponding to $\widetilde\phi$
evolves as
\begin{equation}
\rho\propto A^2\propto R^{-3} \label{rho2}
\end{equation}
where in the last proportionality we have used the solution of
(\ref{eqA2}). This corresponds to the behavior of non-relativistic
matter in the expanding universe.

The second potential to be solved is
\begin{equation}
V=\frac12 m^2\phi^4. \label{pot_quart}
\end{equation}
We shall work in complete analogy with the previous example. We
first find the solution to (\ref{evoleq}) for $H=0$, which is
\begin{equation}
\phi(t)=B\,{\rm sn}(Bmt+\varphi) \label{solphi4H0}
\end{equation}
where sn is the Jacobi elliptic function sn$(Bmt+\varphi,-1)$
\cite{gradshteyn}.

For $H\neq 0$ we choose an ansatz of the form
\begin{equation}
\widetilde\phi(t)=B(t){\rm sn}(B(t)\lambda mt+\varphi).
\label{solphi4H1}
\end{equation}
The calculus here is more complicated than in the previous
example, when introducing (\ref{solphi4H1}) in (\ref{evoleq}). We
obtain a differential equation for the evolution of the amplitude
$B$:
\begin{equation}
\frac{d^2(Bt)^3}{dt^2}+\left(3H-\frac2t\right)\frac{d(Bt)^3}{dt}=0.
\label{eqB4}
\end{equation}
This equation can be easily solved for the usual form
$H=c\,t^{-1}$, where the value of the constant $c$ depends on
which kind of energy dominates the universe ($c=1/2$ for a
radiation-dominated universe and $c=2/3$ for a matter-dominated
one). From (\ref{eqB4}) we obtain
\begin{equation}
\frac{d(Bt)^3}{dt}\propto t^{2-3c}
\end{equation}
that gives
\begin{equation}
B\propto t^{-c}.\label{Bt4}
\end{equation}
Since we are assuming $H=c t^{-1}$, which gives $R\propto t^c$, we
conclude that, for any value of $c$,
\begin{equation}
B\propto R^{-1}. \label{B4}
\end{equation}
Also, we get an equation for $\lambda$, which now depends on the
relative variation of $B$
\begin{equation}
\lambda=\left(1+\frac{\dot B}{B}t\right)^{-1}
=\frac1{1-c}\label{lambda4}
\end{equation}
where in the last equality we took into account (\ref{Bt4}).

The energy density in this case is given by
\begin{equation}
\rho=\frac12(\dot{\widetilde\phi})^2+\frac12m^2(\widetilde\phi)^4
=\frac12m^2B^4
\end{equation}
where we have used properties of the Jacobi elliptic functions
\cite{gradshteyn}. Thus, taking into account (\ref{B4}), we have
\begin{equation}
\rho\propto B^4\propto R^{-4}. \label{rho4}
\end{equation}
This type of $R$-dependence corresponds to a fluid of relativistic
particles, or equivalently, to radiation. The form of the
oscillations in a $\phi^4$ potential with small friction $H$ was
also found in \cite{Greene:1997fu}, in the context of preheating
after inflation. The method used in \cite{Greene:1997fu} was to
make a conformal transformation of the space-time metric and the
fields, as well as a number of re-scalings. Even if we get the
same final results, we present our method because it is somewhat
simpler.

The two examples discussed here are particular cases in which it
is possible to analytically solve the equation of motion that
describes the oscillations of the field $\phi$ in the expanding
universe. For other potentials that can be considered, it might be
impossible to find analytical solutions to (\ref{evoleq}), so the
method applied before might not succeed in obtaining the
dependence of the energy density $\rho$ on the scale factor $R$.
For this purpose, one needs to find another method, in which the
evolution of $\rho(R)$ can be obtained without necessity of
solving the equation of motion (\ref{evoleq}). This is what we
develop in the next section.

\section{Action-angle formalism}\label{section3}

We are concerned with the oscillations of the $\phi$-field about
some minimum of the potential, but we shall {\it not} restrict to
the case that the oscillation amplitude is small. Thus, we are
faced with a system with a periodic motion, whose details might be
complicated. Often, we are not specially interested in these
details, but rather in the frequencies of the oscillations, the
contribution to the energy density, etc. A method of analytical
mechanics tailored for such a situation is provided by the use of
action-angle variables.

As before, it is convenient to start with $H=0$ in (\ref{evoleq});
we have then a conservative system with hamiltonian density $\cal
H$
\begin{equation}
{\cal H}(\phi, \Pi)=\frac12\Pi^2+V(\phi)\equiv\rho
\label{hamiltonian}
\end{equation}
with the momentum $\Pi=\dot\phi$ and energy density $\rho$. When
having a periodic motion, one introduces \cite{goldstein} the
action variable
\begin{equation}
J\equiv\oint\Pi\,d\phi \label{defin_J}
\end{equation}
where the integration is over a complete period of oscillation.
Using $\Pi=\sqrt{2(\rho-V)}$ from (\ref{hamiltonian}), $J$ can be
written as
\begin{equation}
J=2\int_{\phi_{\rm min}}^{\phi_{\rm max}}\sqrt{2(\rho-V)}d\phi
\label{expr_J}
\end{equation}
where $\phi_{min}$ and $\phi_{max}$ are the return points,
$V(\phi_{min})=V(\phi_{max})=\rho$. $J$ is chosen as the new
(conserved) momentum in the integration of the Hamilton-Jacobi
equation. The generating function given by the abbreviated action
\begin{equation}
W=\int\Pi d\phi \label{abb_action}
\end{equation}
allows to canonically transform $(\phi, \Pi)$ into $(\alpha, J)$,
with the angle variable defined by
\begin{equation}
\alpha=\frac{\partial W}{\partial J}.\label{def_alpha}
\end{equation}
Since $W$ does not explicitly depend on time, the new hamiltonian
$\bar{\cal H}$ coincides with the old one $\cal H=\rho$. The
energy $\rho$ is only a function of $J$, which amounts to say that
$\alpha$ is cyclic and $J$ is constant
\begin{equation}
\dot{J}=-\frac{\partial\bar{\cal H}}{\partial \alpha}=0.
\end{equation}
The other Hamilton equation is
\begin{equation}
\dot{\alpha}=\frac{\partial\bar{\cal H}(J)}{\partial J}\equiv\nu,
\label{def_nu}
\end{equation}
with $\nu=\nu(J)$ constant. We can integrate (\ref{def_nu}) to
obtain
\begin{equation}
\alpha(t)=\nu t+\alpha(0) \label{eq_alpha}
\end{equation}
and in a complete period, $\alpha(\tau)=\alpha(0)$. In this way,
we identify $\nu$ in (\ref{def_nu}) with the frequency of the
motion:
\begin{equation}
\nu=\frac{1}{\tau}=\frac{d\rho}{dJ}. \label{frequency}
\end{equation}
Until here, we have reminded ourselves of the standard approach
that uses action-angle variables to find the frequency
\cite{goldstein}.

The realistic case of the expanding universe, with $H\neq 0$, can
be treated with the same method, provided $H$ and $\dot H/H$ are
small compared to the frequency of the $\phi$ oscillations, that
is to say,
\begin{equation}
H\ll\nu\,; \;\; \dot{H}/H\ll\nu. \label{approx}
\end{equation}
Notice that for the usual form $H=ct^{-1}$, $\dot H/H \sim H$ so
that the two conditions in (\ref{approx}) are actually the same.
The conditions (\ref{approx}) ensure that the motion is almost
periodic, and that it makes sense averaging over one cycle. The
$H$-term  in (\ref{evoleq}) represents a time-dependent friction,
so we expect energy to decrease; however, we are assuming this
friction small enough so that we can consider the energy constant
in one-cycle period.

We still define $J$ as in (\ref{defin_J}) and (\ref{expr_J}), and
also the energy density $\rho$ as in (\ref{hamiltonian}). The
decrease rate of $\rho$ can be obtained rewriting the equation of
motion (\ref{evoleq}),
\begin{equation}
\frac{d\rho}{dt}=-3H\dot{\phi}^2. \label{eq_energy}
\end{equation}
When averaging the r.h.s. over one cycle, we can take $H$ out of
the average, $\langle H\dot\phi^2\rangle\simeq
H\langle\dot\phi^2\rangle$. The average of $\dot{\phi}^2$ is
related to $J$:
\begin{equation}\begin{array}{ccl}
\langle\dot{\phi}^2\rangle & = &
\frac{1}{\tau}\oint\,\Pi\dot\phi\, dt\\
& = & \frac{1}{\tau}\oint\Pi d\phi=\frac{1}{\tau}J
\end{array}
\end{equation}
(this is essentially the virial theorem). Thus,
\begin{equation}
\langle\dot\phi^2\rangle =\frac{d\rho}{dJ}J. \label{aver_phi2}
\end{equation}
With the help of (\ref{aver_phi2}), (\ref{eq_energy}) reads
\begin{equation}
\frac{1}{J}\frac{dJ}{dt}=-3H. \label{J_evol}
\end{equation}
What is remarkable about this equation is that the change in the
action variable $J$ is {\it independent} of the form and details
of $V$. The equation can be readily integrated to show that $J$
dilutes as a volume in an expanding universe
\begin{equation}
J\propto R^{-3}. \label{evol_J}
\end{equation}
Eqs.(\ref{J_evol}) and (\ref{evol_J}) are one of our main results.
To find more, we notice that the pressure can also be averaged in
one oscillation period,
\begin{equation}
p=\langle\frac12\dot\phi^2\rangle -\langle V\rangle
=\langle\dot\phi^2\rangle -\rho=w\rho.
\end{equation}
To calculate $w$, we first use (\ref{aver_phi2}) and get
\begin{equation}
w=\frac{J}{\rho}\frac{1}{dJ/d\rho}-1 \label{form_w}
\end{equation}
and, introducing (\ref{expr_J}), we finally obtain
\begin{equation}
w=\frac{2}{\rho}\frac{\int_{\phi_{\rm min}}^{\phi_{\rm
max}}(\rho-V)^{1/2}d\phi}{\int_{\phi_{\rm min}}^{\phi_{\rm
max}}(\rho-V)^{-1/2}d\phi}-1. \label{formula_w}
\end{equation}
Eq.(\ref{formula_w}) coincides with the corresponding expression
in Ref.\cite{Turner}, where another formalism was used.

There is another useful relation that can be easily obtained using
our formalism. Whenever $w$ is constant, we can integrate
(\ref{form_w}) to obtain
\begin{equation}
\rho\propto J^{w+1}.
\end{equation}
Then, making use of (\ref{evol_J}), we get the well-known relation
\begin{equation}
\rho\propto R^{-3(w+1)}. \label{rho_de_R}
\end{equation}

As an example, let us apply the method described above for the
following power-law potential
\begin{equation}
V=a\phi^n \label{power_law_pot}
\end{equation}
for which the action variable can be exactly calculated
\begin{equation}
J=\frac{4\sqrt{2\pi}\,
\Gamma\left(\frac1n\right)}{(n+2)\Gamma\left(\frac12+\frac1n\right)}
\sqrt{\rho}\left(\frac{\rho}{a}\right)^{1/n}. \label{J_powerlaw}
\end{equation}
Using (\ref{frequency}), we get the frequency of the motion
\begin{equation}
\nu=\frac{n
\Gamma\left(\frac12+\frac1n\right)}{2\sqrt{2\pi}\,\Gamma\left(\frac1n\right)}
\sqrt{\rho}\left(\frac{\rho}{a}\right)^{-1/n}
\end{equation}
and using (\ref{form_w}), we get the parameter $w$
\begin{equation}
w=\frac{n-2}{n+2}. \label{w_aphin}
\end{equation}
Let us show that these results coincide with what we obtained in
Sect.\ref{section2}. For $n=2$, we get from (\ref{w_aphin}) that
$w=0$, which corresponds to the equation of state for
non-relativistic matter, $p=0$. Equivalently, from
eq.(\ref{rho_de_R}), we obtain that
\begin{equation}
\rho\propto R^{-3}
\end{equation}
which is the same as (\ref{rho2}). Still for the harmonic
potential, $n=2$, taking $a=\frac12m^2$, we can calculate $J$ by
making use of (\ref{J_powerlaw})
\begin{equation}
J=2\pi\frac{\rho}m \label{J_energ}
\end{equation}
which is nothing else than the number density of particles of mass
$m$, except the factor of $2\pi$. We now see that, in this case,
(\ref{evol_J}) is equivalent to assert that the number density
decreases as $R^{-3}$ in an expanding universe.

For $n=4$, (\ref{w_aphin}) gives $w=1/3$, and (\ref{rho_de_R})
gives $\rho\propto R^{-4}$ which coincides with (\ref{rho4}) and
corresponds to the equation of state for relativistic matter.

As expected, we have recovered the same results of
Sect.\ref{section2} without the need to solve the equation of
motion, eq(\ref{evoleq}). For this reason, the action-angle
formalism will allow us to study more general cases of potentials
for which there is no analytical solution to (\ref{evoleq}). This
is what we will do in the next section.

As another application of the action-angle variables, we discuss
the issue of adiabatic invariants. When a parameter $a$ of the
potential changes with time slowly compared with the natural
frequency of the system,
\begin{equation}
\frac{\dot a}{a}\ll\nu, \label{param_var}
\end{equation}
we say it changes adiabatically. When there is such an explicit
time change in the potential, $W$ in (\ref{abb_action}) depends on
time, and $\bar{\cal H}$ and $\rho$ do not coincide any longer
\begin{equation}
\bar{\cal H}=\rho+\frac{\partial W}{\partial t}=\rho+W_{,a}\dot a
\end{equation}
where the derivative $W_{,a}\equiv\partial W/\partial a$ is taken
at constant $J$. Instead of (\ref{J_evol}), we now have
\begin{equation}\begin{array}{ccl}
\dot J & = & -3HJ-
\langle\partial \bar{\cal H}/\partial\alpha\rangle \\
& = & -3HJ- \langle (\partial W_{,a}/\partial\alpha)\dot a\rangle.
\end{array}
\end{equation}
When averaging this identity, due to (\ref{param_var}), we can
consider $\dot a$ as constant during a cycle, i.e., $\langle
(\partial W_{,a}/\partial\alpha)\dot a\rangle\simeq \langle
(\partial W_{,a}/\partial\alpha)\rangle\dot a$. The average
$\langle\partial W_{,a}/\partial\alpha\rangle$ vanishes, since
$W_{,a}$ is a periodic function of $\alpha$. It follows that
\begin{equation}
\dot J=-3HJ, \label{J_evol2}
\end{equation}
namely, eqs.(\ref{J_evol}) and (\ref{evol_J}) are still valid,
although we stress that $V$ is changing with time. For $H=0$, the
system is conservative and we have $\dot J=0$, i.e., $J$ is
identified with an adiabatic invariant \cite{goldstein}.

An instance where we can apply these results is the invisible
axion model. In this model we have a time-dependent axion mass,
$m=m_a(t)$ in (\ref{pot_quadrat}). Even if $\rho(t)$ and $m_a(t)$
are complicated functions of time, we conclude after our study
that, provided the change is adiabatic, the combination
\begin{equation}
\frac{J}{2\pi}=\frac{\rho(t)}{m_a(t)}\propto R^{-3}
\label{J_axion}
\end{equation}
behaves as the axion number density. This behavior is what is
expected on general grounds. This result was derived in
\cite{threepapers} when calculating the relic axion density. What
we have done in the present paper is a kind of generalization of
(\ref{J_axion}), that may be applied when having other potentials
and/or other types of adiabatic change.

\section{Dark energy from field oscillations}\label{section4}
In the standard model of cosmology, besides eq.(\ref{Fried_eq}),
there is another equation that one can get from Einstein
equations. It can be put in the form
\begin{equation}
\frac{\ddot R}{R}=-\frac{4\pi G}{3}(\rho+3p).
\end{equation}
For the simple equation of state $p=w\rho$, we see that $w<-1/3$
leads to an accelerated universe. We will show in this section
that such values of $w$ can be obtained from an oscillating scalar
field whose energy density dominates. An application based on this
idea in the context of inflation was developed by Damour and Mukhanov
in \cite{damour} (see also \cite{sami}).

In order to check whether a given potential may lead to
acceleration, we will calculate $w$ using the expression
(\ref{form_w}), that involves an average over one cicle
\footnote{In \cite{damour}, a nice geometric interpretation of the
condition $w<-1/3$ is given.}. The value of $w$ depends on the
form of the potential and on the return values $\phi_{max}$ and
$\phi_{min}$. In turn, these change with time due to the friction
produced by the expansion of the universe, so that, in general,
$w$ depends on time.

A remarkable exception to this dependence on the return values is
given by the power-law potential (\ref{power_law_pot}). As shown
in (\ref{w_aphin}), $w$ only depends on the power $n$. For integer
$n$, we do not have oscillations neither for $n=0$, nor for $n=1$.
The potential $V=a\phi^2$ leads to oscillations that correspond to
non-relativistic matter with $w=0$. The next potential leading to
oscillations is $V=a\phi^4$, that we saw it implies $w=1/3$.
Higher values of the power $n$ lead to higher values for $w$. No
integer value of $n$, leading to oscillations, can correspond to a
fluid with $w<-1/3$ \footnote{The case of a non-integer $n<2$ is
considered in \cite{hsu}. Power law potentials with scaling
solutions are studied in \cite{scherrer}.}.

In order to have smaller $w$ values, one clearly needs a field
$\phi$ that, even oscillating, has a slow-roll for enough time. To
illustrate that this is possible, we shall investigate a few
potentials. In all of them, we shall put the center of the
oscillations at $\phi=0$. In addition, we will work with symmetric
potentials, so that $-\phi_{min}=\phi_{max}\equiv \phi_0$.

\begin{figure}[htb]
\begin{center}
\includegraphics[width=7cm, height=5cm]{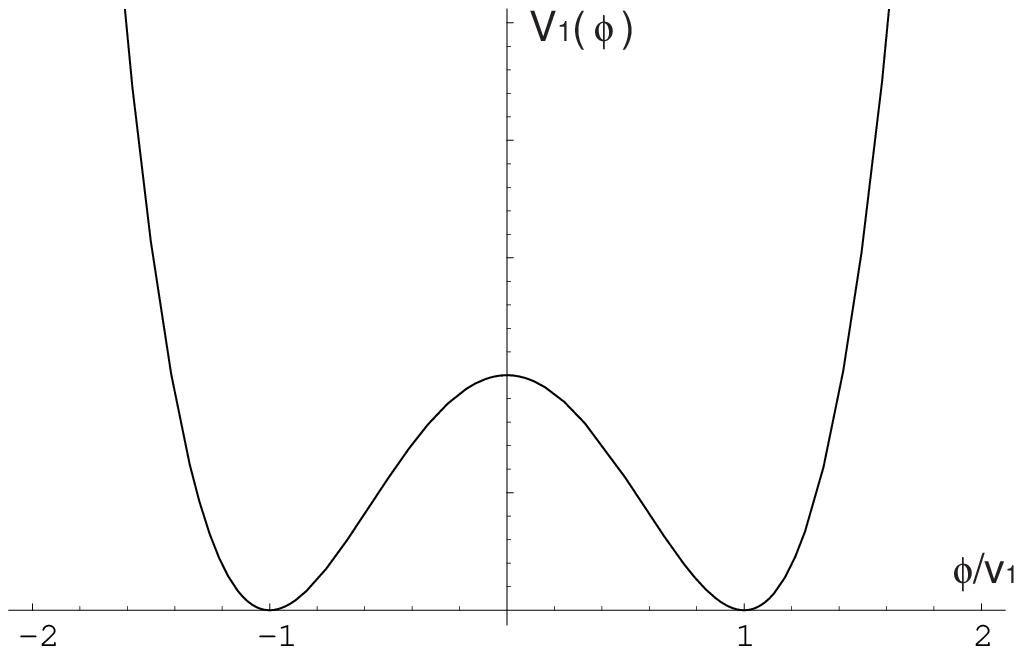}
\end{center}
\caption{"Mexican hat" potential $V_1$ defined in
(\ref{potV1}).}\label{fig1}
\end{figure}

\begin{figure}[htb]
\begin{center}
\includegraphics[width=7cm, height=5cm]{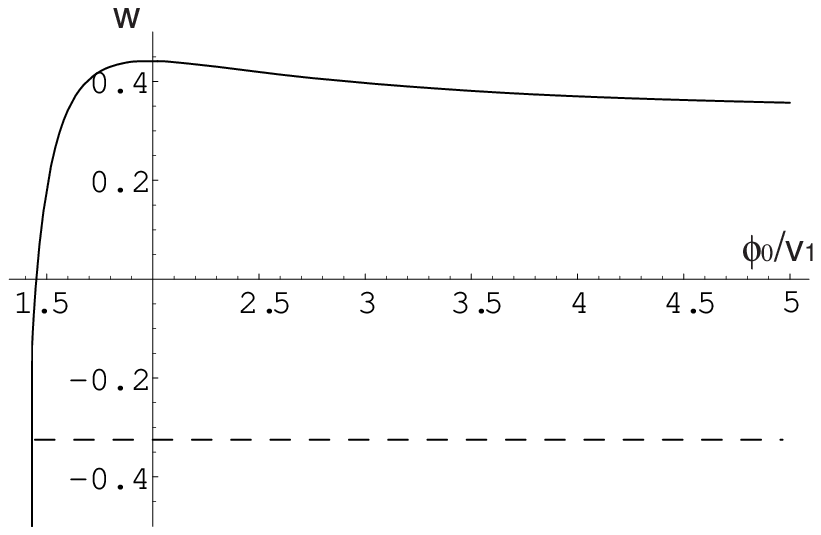}
\end{center}
\caption{$w$ as a function of the amplitude $\phi_0$ of the
oscillations in the potential $V_1$. The dashed line indicates
$w<-1/3$.} \label{fig2}
\end{figure}

Let us start with the "Mexican hat" potential shown in
Fig.{\ref{fig1}},
\begin{equation}
V_1(\phi)=\lambda_1(\phi^2-v_1^2)^2 \label{potV1}
\end{equation}
with $\lambda_1$ a coupling and $v_1$ an energy scale. We consider
$\phi_0>\sqrt2 v_1$, to have oscillations around $\phi=0$. The
parameter $w$, calculated numerically, is a function of
$\phi_0/v_1$ and it is displayed in Fig.\ref{fig2}. We see that
when $\phi_0$ is close enough to $\sqrt2 v_1$, we have values
$w<-1/3$. However, an unsatisfactory feature is that one has to
tune the return value $\phi_0$.

\begin{figure}[htb]
\begin{center}
\includegraphics[width=7cm, height=5cm]{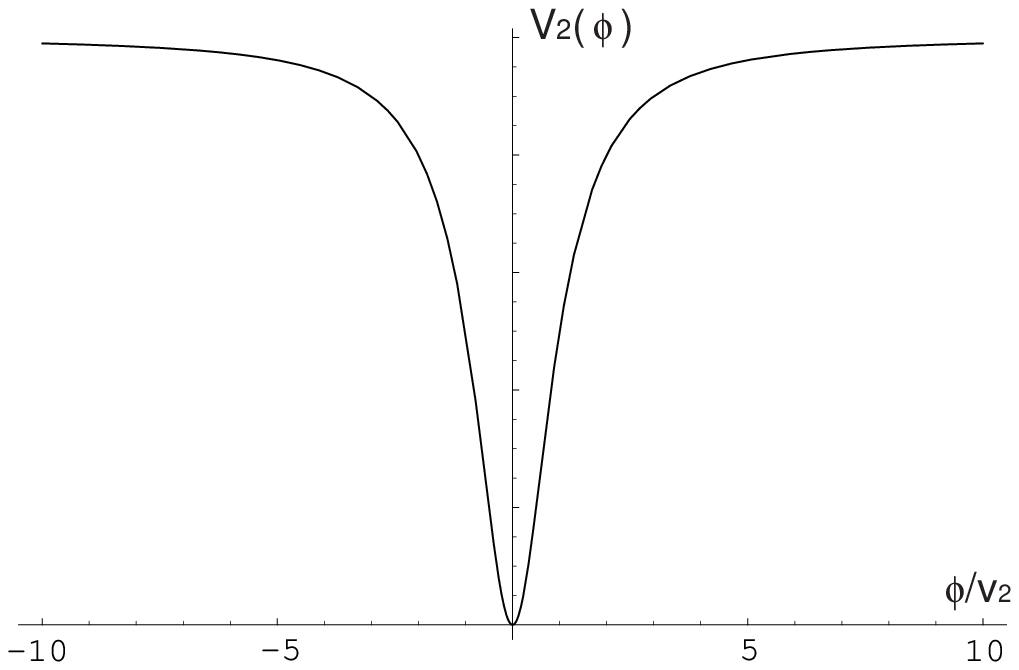}
\end{center}
\caption{The shape of the potential $V_2$ defined in
(\ref{potV2}).}\label{fig3}
\end{figure}

\begin{figure}[htb]
\begin{center}
\includegraphics[width=7cm, height=5cm]{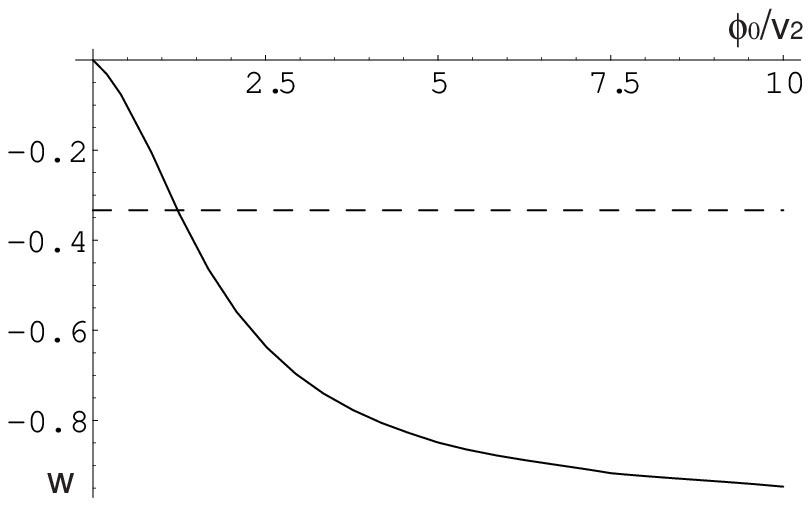}
\end{center}
\caption{$w$ as a function of the amplitude $\phi_0$ of the
oscillations in the potential $V_2$. The dashed line indicates
$w<-1/3$.} \label{fig4}
\end{figure}

A potential that does not have this fine-tuning problem is given
by
\begin{equation}
V_2(\phi)=\frac{{v_2'}^4\phi^2}{\phi^2+v_2^2} \label{potV2}
\end{equation}
(see Fig.\ref{fig3}). Here, $v_2$ and $v'_2$ are energy scales.
The values of $w$ are shown in Fig.\ref{fig4}, where we see that
$w<-1/3$ for $\phi_0>1.2\,v_2$. To get the observed acceleration
of the universe, we should have $v'_2\simeq 2\times 10^{-3}$ eV,
as expected. Notice that in the past, field oscillations give
$w\simeq -1$ always. As happens in the case of a cosmological
constant, the contribution we are discussing was subdominant in
the past, as soon as non-relativistic matter enters in the stage
and dominates.

One could find other potentials giving acceleration. One such
example was discussed in \cite{damour}. There, the objective was
to get first slow-roll inflation and, after, oscillations
provoking further inflation. In contrast to \cite{damour}, we work
in the period of dark energy domination and we would like to know
the future of the universe. We see in Fig.\ref{fig4} that the
oscillations will end up giving $w=0$, i.e., being harmonic and
consequently matter dominated.

\begin{figure}[htb]
\begin{center}
\includegraphics[width=7cm, height=5cm]{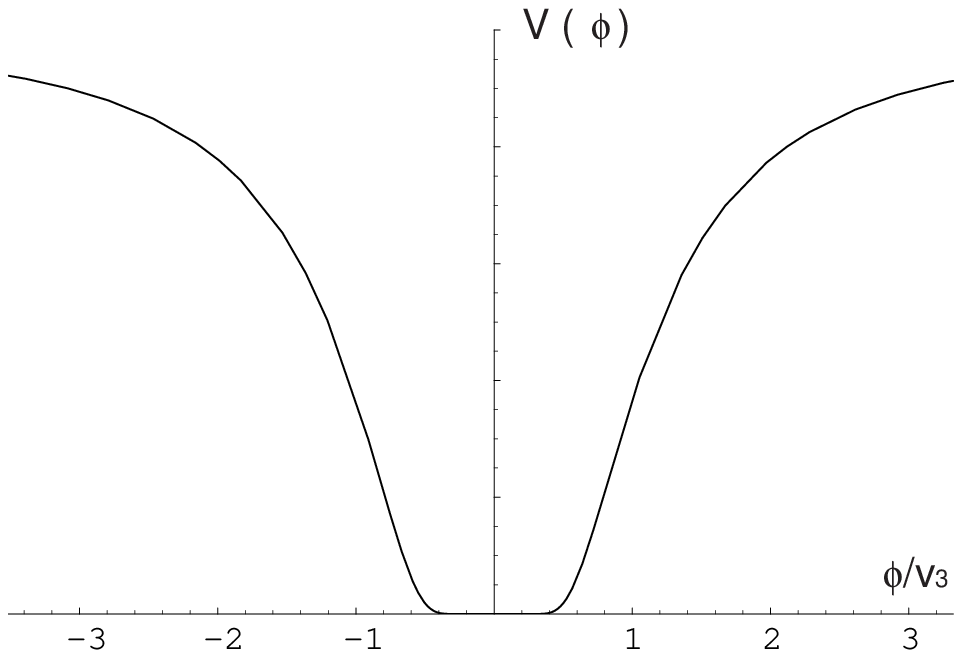}
\end{center}
\caption{The shape of the potential $V_3$ defined in
(\ref{pot_exp}).}\label{fig5}
\end{figure}

\begin{figure}[htb]
\begin{center}
\includegraphics[width=7cm, height=5cm]{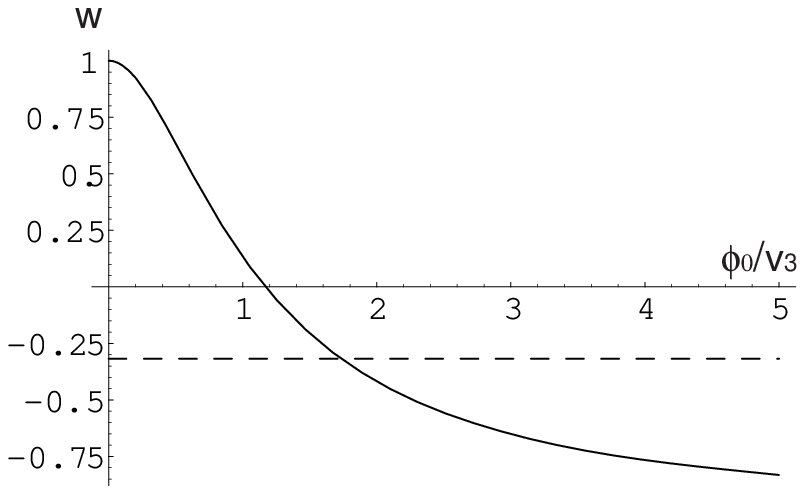}
\end{center}
\caption{$w$ as a function of the amplitude $\phi_0$ of the
oscillations in the potential $V_3$. The dashed line indicates
$w<-1/3$.}\label{fig6}
\end{figure}

One may think that for all potentials, when the field is close
enough to the minimum, one ends with the usual harmonic potential
(\ref{pot_quadrat}) that well approximates oscillations around the
minimum, so that one obtains $w=0$. We would
like to point out that there is at least one exception. Consider
the potential
\begin{equation}
V_3(\phi)={v_3'}^4 e^{-v_3^2/\phi^2} \label{pot_exp}
\end{equation}
with $v_3$ and $v'_3$ some energy scales (see Fig.\ref{fig5}). An
inspection to the potential shows that for $\phi_0$ large enough,
we should have $w<-1/3$. We have calculated $w$ and our results
are shown in Fig.{\ref{fig6}}. Indeed, we see that $w<-1/3$ for
$\phi_0>1.8\,v_3$. For smaller values of $\phi_0$, higher values
of $w$ are obtained. We notice that in the limit
$\phi_0/v_3\rightarrow 0$ the value $w=0$ is not obtained, but
rather $w=1$. This corresponds to a fluid with $p=\rho$. The
reason for this very particular behavior is the well-known fact
that $V_3(\phi)$ in (\ref{pot_exp}) can not be developed in a
Taylor-Maclaurin series around $\phi=0$. Again we should check
that $V_3$ is consistent with the observational constraints. With
$v'_3\simeq 2\times 10^{-3}$ eV the oscillations fit the
acceleration. Also, for early times $w\simeq -1$ so that the
oscillations do not contradict the observed evolution of the
universe.

Let us finally address the requirement that dark energy should
have very homogeneous density, with spatial irregularities
rearranging at an effective relativistic speed of sound $c_s$,
given by
\begin{equation}
c_s^2=\frac{dp}{d\rho} \label{sound_speed}
\end{equation}
In the main spirit of our paper, we should point out that
(\ref{sound_speed}) can be calculated following the technique of
action-angle variables. This allows to find $c_s$ without
reference to the details of the solution to the equation of motion
of the field. The way we do it is through the equation
\begin{equation}
c_s^2=\frac{d(w\rho)}{d\rho}=\frac{\rho
dw/d\phi_0}{d\rho/d\phi_0}+w= \frac{V_0 dw/d\phi_0}{dV_0/d\phi_0}
+w \label{sound_speed2}
\end{equation}
where $V_0\equiv V(\phi_0)=\rho$ and $w$ was found in
(\ref{form_w}) or (\ref{formula_w}).

As a check, we have calculated the integrals (\ref{sound_speed2})
for the potential (\ref{pot_quadrat}) and obtained $c_s^2\simeq
0$, as expected. The implications of a scalar field with the
potential (\ref{pot_quadrat}) when it dominates the energy density
of the universe have been fully investigated in ref.\cite{ratra}.
Here we do not pretend to do such a complete study for the
potentials of Sect.\ref{section4} since our emphasis is in the
techniques developed in Sect.\ref{section3}. Let us point out,
however, that in fact we find that for the potentials
(\ref{potV1}), (\ref{potV2}) and (\ref{pot_exp}), $c_s^2$ is near
zero or even negative, which indicates that collapse at small
scales would happen or even would lead to dangerous instabilities.

It is not difficult, however, to find potentials giving $w\leq
-1/3$ and $c_s^2\simeq 1$. One example is
\begin{equation}
V_4(\phi)=v_4'^4\frac{\phi^2}{v_4^2+\phi^2}+ \frac14\lambda\phi^4
\label{potV4}
\end{equation}
We have verified that $w$ from (\ref{formula_w}) and $c_s^2$ from
(\ref{sound_speed2}) for the potential $V_4$ in (\ref{potV4}) have
the expected values for a dark energy contribution, for a suitable
range for the potential free parameters.


\newpage

\end{document}